\documentclass{IEEEtran}
\usepackage{cite}
\usepackage{amsmath,amssymb,amsfonts}
\usepackage{algorithmic}
\usepackage{graphicx}
\usepackage{textcomp}
\def\BibTeX{{\rm B\kern-.05em{\sc i\kern-.025em b}\kern-.08em
    T\kern-.1667em\lower.7ex\hbox{E}\kern-.125emX}}

\usepackage{epsfig}
\usepackage{xcolor}
\usepackage{float}
\usepackage{graphicx}
\usepackage{cite}
\usepackage{tikz}
\usepackage{booktabs}

\usepackage{colortbl}

\newcommand{\rone}[1]{\boldsymbol{#1}}
\newcommand{\rtwo}[1]{\underline{\underline{\boldsymbol{#1}}}}

\definecolor{Gray}{gray}{0.85}
\definecolor{LightCyan}{rgb}{0.88,1,1}

\usepackage{array}
\usepackage{makecell}
\usepackage{rotating}

\usepackage[super]{nth}

\ifCLASSINFOpdf
\else
\fi
\ifCLASSOPTIONcompsoc
 \usepackage[caption=false,font=normalsize,labelfont=sf,textfont=sf]{subfig}
\else
 \usepackage[caption=false,font=footnotesize]{subfig}
\fi

\begin{document}


\title{{Analytical Model For The Contribution Of Small Scatterers to Open-Ended Coaxial Probe Measurements}}

\author{Rotem Gal-Katzir,~\IEEEmembership{Student Member,~IEEE,}
        Emily~Porter,~\IEEEmembership{ Member,~IEEE,}
        Yarden~Mazor,~\IEEEmembership{Senior Member,~IEEE,}
\thanks{R. Gal-Katzir and Y. Mazor The authors are with the department of electrical engineering in Tel-Aviv University, Israel, 66978}
\thanks{E. Porter is with the Department of Biomedical Engineering, McGill University, Montreal, Canada, H4A 3J1}
}

\markboth{ }%
{Shell \MakeLowercase{\textit{et al.}}: Bare Demo of IEEEtran.cls for IEEE Journals}
%

\date{\vspace{-5ex}}
\maketitle

\begin{abstract}
The open-ended coaxial probe (OECP) technique is one of the most commonly used methods for the characterization of homogeneous media properties, especially in the biomedical sciences. However, when considering inhomogeneous media, the effect of the heterogeneity on the probe terminal admittance is unclear, making the measured admittance hard to interpret and relate to the medium properties. In this paper we present an analytical model for the contribution of an isotropic scatterer embedded in an otherwise homogeneous medium to the probe admittance. We utilize rigorous scattering theory and various approximations to obtain simplified, closed-form expressions. Using the obtained results we present a method to accurately extract the scatterer properties from a measurement of the admittance. In addition, we define the sensing depth, and show how it can be mapped as a function of the expected scatterer properties. Full-wave simulations are used to verify the analytical model, and the proposed method paves a path for further generalization to additional scenarios of open-coaxial probe sensing of an inhomogeneous medium. 

\end{abstract}


\IEEEpeerreviewmaketitle

%
%
%

\section{Introduction}

\IEEEPARstart{O}{ne} of the simplest and most simple and common methods for characterizing dielectric properties across the RF/MW range, is the open-ended coaxial probe (OECP) technique. It is the most commonly and widely used today to characterize liquids, semi-solids and solids, including biological tissues \cite{Mosig},\cite{Sarolic}. This technique has found particular utility due to its non-destructive nature (samples do not need to be shaped or prepared in any specific way), the ability to measure in-vivo or ex-vivo, and ease in performing broadband measurements across wide temperature ranges \cite{Diagnostics}.

One of the critical challenges in deriving the dielectric properties of biological tissues using an OECP is that the common extraction algorithms rely on the assumption that the medium is homogeneous. On top of that, most strategies used to deal with actual sample heterogeneities lack a fundamental basis in wave theory and have been shown to lead to highly inconsistent and error-prone results \cite{GioiaInvestigation},\cite{LaGioia}. To enhance the applicability of this method, and to possibly incorporate it in future material science and biomedical technologies is required. In the study of biological samples, several studies have used histology to identify the tissue types and distributions present in a sample that result in measured bulk dielectric properties \cite{Lazebnik,sugitani2014complex}; however, it is unclear how tissues at different locations in the sample contribute to this bulk measurement, even for relatively simple samples \cite{Farshkaran,LaGioia}. 

In this work, we focus on characterizing simple, but inhomogeneous, samples using an OECP probe. 
We present a comprehensive process for the characterization of a small isotropic scatterer within a homogeneous medium. This is achieved first by analytically modeling the inhomogeneity using rigorous scattering theory and approximate models. We use the basic formulation presented in \cite{Papas},\cite{Levine}, and extend it to account for electromagnetic fields scattered by the scatterer. We study the contribution of the scatterer to the normalized terminal admittance seen at the coax-sample interface. Using this, we derive an approximate formulation, from which we estimate the scatterer properties. The scenarios studied are versatile and include parametric studies of the scatterer and backgrounds' permittivities for different scatterer sizes and locations. 
 Finally, we use the analytical results to map the sensing depth of the OECP technique for various types of probes and materials as a function of the scatterer properties. This aspect is particularly important when envisioning the extension of the OECP technique to inhomogeneous media, as it determines what region, and thus which tissues within the sample, will contribute to the dielectric measurement \cite{Emily_2018}, and gives us a tool to classify whether or not we can expect to detect the inhomogeneity.
Throughout our work, we compare our results against full-wave simulations performed in COMSOL \cite{COMSOLref}, as a benchmark to evaluate the accuracy of the derived formulae and the possible error sources.
The presented model demonstrates highly consistent results and provides insights into the physics of the setup, further enhancing our understanding of how heterogeneities affect the properties of biological tissues.  



This article is organized as follows: the first subsection in Section \ref{AnalyticalModelSection}, discusses notations and the configuration at hand. In the following subsections the analytical model for the contribution of the scatterer to the measurement is elaborated.
Then, in Section \ref{Applications} we present the results compared with COMSOL simulations. In this section, we propose two implementations of the derived analytical model - parameter estimation and a novel approach for the calculation of the sensing depth. Finally, in Section \ref{Conc}, concluding remarks are presented.

\section{Analytical Model for a Localized Scatterer}\label{AnalyticalModelSection}

\subsection{Notation and Configuration}\label{sec:ProblemConf}
Unless stated otherwise for a specific parameter, we use the following notation conventions: an italic font is used for scalars (e.g. $a , A$) and bold font is used for vectors (e.g. ${\bf a, A}$).

\begin{figure}[ht]
\begin{center}
\noindent
  \includegraphics[width=3.2in]{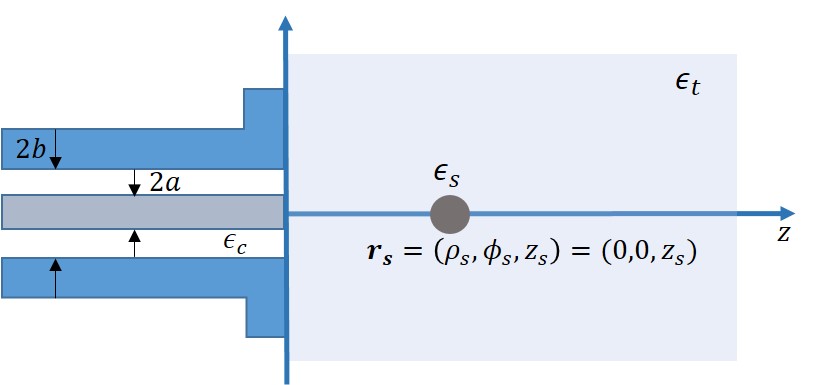}
\caption{The OECP and isotropic scatterer inhomogeneity configuration}
\label{ConfigurationFig}
\end{center}
\end{figure}

We consider the case of a small dielectric isotropic scatterer, inserted in a homogeneous background medium, and positioned in the axial direction of an open-ended coaxial probe.
The configuration and its parameters are shown in Fig. \ref{ConfigurationFig}. The scatterer complex permittivity is $\epsilon_s = \epsilon_s'+i \epsilon_s''$, where $\epsilon_s''=\frac{\sigma}{\omega \epsilon_0}$, with $\sigma$ being the conductivity, $\omega$ is the angular frequency and $\epsilon_0$ is the vacuum permittivity. Throughout this work, the term "permittivity" will be used to indicate complex permittivity. The scatterer is located on the z-axis, hence ${\bf r_s} = (\rho_s , \phi_s ,z_s) = (0 , 0 , z_s)$, where the subscript $(\_)_s$ stands for “scatterer”. $\epsilon_{t}, \mu_{t}$ are respectively the permittivity and permeability of the test sample surrounding the scatterer. The test sample is in contact with an open-ended coaxial cable, with a perfect electric conductor (PEC) flange termination. The inner and outer radii of the coaxial line are specified by $a$ and $b$, respectively, and its insulator permittivity is given by $\epsilon_c$. For consistency with previous related works, the time dependence is taken to be $e^ {-i\omega t}$ \cite{Papas}.

\subsection{Dipole Moment Contribution}\label{sec:AnalyticalModelDipole}
To derive the contribution of the small scatterer to the terminal admittance, we first need to describe its contribution to the coaxial aperture field through a new modified Green's function for the half-space $z>0$. To describe the fields of the scatterer, we use the multipole expansion \cite{papas1965theory,Alu,Quad_Alu}. Here we assume that $\bf{E(r_s)}$ is the local electric field applied on the scatterer, located at ${\bf r_s}$. 

If we expand the impinging field distribution in Taylor series around ${\bf r_s}$, we may derive a series expression for scattered field due to electric multipoles 

\begin{equation}
{\bf E_{scat}(r)} = \left(\rone{p}+\frac{1}{2}\rtwo{Q}\cdot\nabla+...\right)\rtwo{G}
\label{Background : Multipole Scat Field}
\end{equation}
with $\rtwo{G}$ being the Green's function.
${\bf p}$ and ${\bf Q}$ represent the electric dipole moment and electric quadrupole moment, respectively. For a small particle, the electric-dipole moment  ${\bf p_s}$ is dominant, with an amplitude proportional to the local electric field
\begin{eqnarray}
\nonumber {\bf p_s} &=& \alpha_d \bf{E(r_s)} \\ &=& -\frac{4 \pi \epsilon_{t}}{k^3} a_1
\bf{E(r_s)} 
\label{Formulation : Ps}
\end{eqnarray}
where $\alpha_d$ is the electric polarizability (denoted here by a scalar due to isotropy of the scatterer), $a_1$ is the first Mie scattering coefficient \cite{Kong},  $\epsilon_t$ is the permittivity of the background material and $k$ is the wave-number satisfies $k^2 = \omega^2 \epsilon_{t} \mu_{t}$.

Then, the new modified Green's function can be derived by considering the magnetic field of a dipole in free-space \cite{Kong} 
\begin{eqnarray}
{\bf H}_{dip}(\bf r) &=& - i\omega 
\left( -ik + \frac{1}{r}\right)
\left( {\bf p_s \times \hat{r}}\right)
\frac{e^{ikr}}{4 \pi r}
\label{Formulation : scatterer H}
\end{eqnarray}

With the given parameters, we can consider the scatterer to be in a deep-subwavelength regime. Therefore, the coefficient $a_1$ can be reliably expressed by the quasi-static approximation, under which the polarizability $\alpha_d$ reduces to \cite{Tretyakov}
\begin{eqnarray}
\alpha_d ^ {-1} &=& 
\frac{1}{4\pi \epsilon_{t} a_s^3}  \frac{\epsilon_{s} + 2 \epsilon_{t}} {\epsilon_{s} - \epsilon_{t}}
-i \frac{k^3}{6 \pi \epsilon_{t}}
\label{Formulation : alpha approc}
\end{eqnarray}

Here $a_s$ is the radius of the scatterer. Note that this result also includes the effect of radiation loss, rendering $\alpha_d$ a complex number. 

To derive the dipole moment of the scatterer, we need to determine the local field applied at ${\bf r_s}$. We calculate the fields outside the coax by expressing the equivalent sources in the aperture plane, and using the modified Green's function for a half-space ${\bf G}_{ap}({\bf r})$, which takes into account the PEC flange. Due to the PEC termination at $z=0$, an image scatterer with dipole moment ${\bf p_s }$ is added at ${\bf r'_s}= (\rho'_s , \phi'_s , z'_s) = (0,0,-z_s)$. As a result, the scatterer is subjected to the following total electric field 
\begin{eqnarray}
{\bf E(r_s) } &=& 
{\bf E}_{ap}({\bf r_s}) + {\bf p_s } {\bf G}_{dip}^E({\bf r_s - r'_s})
\label{Formulation : E external}
\end{eqnarray}

Here, the subscript $(\_)_{ap}$ stands for “aperture”. ${\bf G}_{dip}^E({\bf r_s - r'_s})$ is the free-space electric dyadic Green's function. On the z-axis, only the z-component of the electric field in the half-space $z>0$ remains, therefore the only relevant component of the electric Green's dyadic is $G^E_{dip,zz}(2z_s)$ -- the zz component, given by
\begin{eqnarray}
 G^E_{dip,zz}(2z_s) &=& 
\left( -ik +\frac{1}{2z_s} \right )
\frac{e^{ik2z_s}}{8\pi\epsilon_t z_s^2}
\label{Formulation : G^E_zz}
\end{eqnarray}
and after using equations (\ref{Formulation : Ps}) and (\ref{Formulation : E external}) we obtain
\begin{eqnarray}
{\bf p_s^{eff}} =
\frac{\alpha_d}{1-\alpha_d G^E_{dip,zz}(2z_s)}
{\bf E}_{ap,z}({\bf r_s})
\label{Formulation :P_s effective}
\end{eqnarray}
$ {\bf E}_{ap,z}({\bf r_s})$ is the z component of the aperture field outside the coaxial probe, that can be expressed using the field in the aperture plane as  
\begin{multline}
{\bf E}_{ap,z}({\bf r_s})  = 
\frac{\hat{z}}{2\pi} 
\int_a^b \varepsilon (\rho ') \rho' d\rho' \\
\times
\int_0^{\pi} d\phi' \rho ' \cos ^2\phi' 
\frac{ e^{ik\sqrt{\rho'^2 +z_s ^2}}}
{\rho'^2 +z_s ^2}
\left(
- ik + \frac{1}{\sqrt{\rho'^2 +z_s ^2}}\right) 
\end{multline}
Here, $\varepsilon (\rho)$ is the $\hat{\rho}$ component of the electric field in the aperture plane. 
With this analysis complete, we are able to obtain the magnetic field in the half space $z>0$ 
\begin{eqnarray}
{\bf H }_{z>0}({\bf r}) &=& 
{\bf H}_\phi^{(+)}({\bf r}) +
{\bf H}_s({\bf r - r_s}) + {\bf H}_s({\bf r + r_s})  
\label{Formulation : H tot}
\end{eqnarray}

${\bf H}_\phi^{(+)}({\bf r})$ is the magnetic field generated by the coaxial aperture, and given in eq. (2.25) in \cite{Papas}. ${\bf H}_s({\bf r \mp{r}_s})$ is the magnetic field radiated by the scatterer and its image, respectively, and it is given by substituting (\ref{Formulation :P_s effective}) into (\ref{Formulation : scatterer H})
\begin{multline}
{\bf H}_s({\bf r \mp rs}) =
{\bf G}^H_{s} ({\bf r \mp r_s})
\frac{\alpha_d}{1-\alpha_d G^E_{dip,zz}(2z_s)}\\ 
\times  
\int_a^b \rho'^2 \varepsilon (\rho ')
\frac{ e^{ik\sqrt{\rho'^2 +z_s ^2}} }{\rho'^2 +z_s ^2} 
\left( -ik + \frac{1}{\sqrt{\rho'^2 +z_s ^2}}\right) d\rho'
\label{Formulation : H sphere}
\end{multline}
%
%
%
where $({\bf r \mp r_s}) = (\rho , \phi , z \mp z_s)$, and ${\bf G}^H_{s} ({\bf r \mp r_s})$ given in cylindrical coordinates is
\begin{multline}
{\bf G}^H_{s} ({\bf r \mp r_s}) =
\frac{i\omega }{4\pi} 
\frac{\rho}{\rho^2 + (z-z_s)^2} 
\\ \times
\left( ik - \frac{1}{\sqrt{\rho^2 + (z-z_s)^2}}\right)
e^{ik\sqrt{\rho^2 + (z-z_s)^2}}
 \hat{\phi}
\label{Formulation : G_H in cyl. Cor.}
\end{multline}

Finally, we account for the scatterer contribution to the fields on the aperture plane of the open ended coaxial probe, in terms of the terminal admittance. To obtain an analytic expression for the terminal admittance, we construct a variational principle following the procedure in \cite{Papas} (see Appendix \ref{AppendixA}). This leads us to the desired representation for the total terminal admittance 
\begin{eqnarray}
\frac{Y_{tot}(0)}{Y_0} &=& 
\frac{Y(0)}{Y_0 } + \frac{Y_p(0)}{Y_0}
\label{Formulation : Y_tot}
\end{eqnarray}

Here $Y_0 $ is the characteristic admittance of the probe, and the scatterer's dipole contribution $\frac{Y_p(0)}{Y_0}$ is provided by
\begin{multline}
\frac{Y_p(0)}{Y_0} = \frac{G_p(0)}{Y_0}
- i \frac{B_p(0)}{Y_0}\\
= -\frac{i\omega }{2\pi} \frac{\eta_c \log\left(\frac{b}{a}\right)}
{[\int_a^b \varepsilon(\rho) d\rho]^2}
\frac{\alpha_d}{1-\alpha_d G^E_{dip,zz}(2z_s) }\\
 \times
 \left[ \int_a^b \varepsilon (\rho) d\rho \frac{\rho^2}{\rho^2 +z_s ^2}
 e^{ik\sqrt{\rho^2 +z_s ^2}} 
 \left( ik - \frac{1}{\sqrt{\rho^2 +z_s ^2}}\right) \right]^2  
\label{Formulation : Y_s}
\end{multline}
%
%
where $\eta_c = \sqrt{\mu_c / \epsilon_c}$ is the intrinsic impedance of the probe insulator. This expression is very versatile, lending itself to various methods of evaluation.  

\subsection{Quadrupole Moment Contribution}\label{sec:AnalyticalModelQuadrupole}
 The secondary contribution of the scatterer to the terminal admittance comes from the electric quadrupole moment ${\bf Q}$, which interacts with the gradient of the external electric field. Using a similar formulation, we define the electric quadrupolarizability of the scatterer given by $\alpha_q$, to obtain the following relation
\begin{eqnarray}
{\bf Q} &=& \alpha_q \frac{{\bf (\nabla E(r_s) + E(r_s) \nabla)}}{2}
\label{Formulation : Q}
\end{eqnarray}
where $\alpha_q$ satisfies \cite{Alu}
\begin{eqnarray}
\alpha_q^{-1} &=& \frac{k^5}{40 \pi \epsilon} 
\left( 15 (ka_s )^{-5} \frac{2\epsilon_s +3\epsilon}{\epsilon_s -\epsilon} -i \right)
\label{Formulation : alpha_q}
\end{eqnarray}

Here, we assume that the external field and its derivative on the symmetry axis is only in the ${\bf \hat{z}}$ direction, therefore we obtain
\begin{eqnarray}
\frac{\partial E_{0z}}{\partial z} &=&
-\frac{1}{i \omega \epsilon_t } 
\left(\frac{\partial }{\partial \rho} + \frac{1}{\rho}\right)
{\bf H }_{z>0}({\bf r})
\label{Formulation : Ez derivative}
\end{eqnarray}
where ${\bf H }_{z>0}({\bf r})$ is defined in (\ref{Formulation : H tot}). Then, the particle's radiation and scattering may be described by the vector potential, which reduces to
\begin{eqnarray}
{\bf A_q(r)} &=& \frac{i \omega \mu_0}{8 \pi} Q_{zz} \frac{z}{r^2 } \left( ik - \frac{1}{r}\right) e^{ikr} {\bf \hat{z}}
\label{Formulation : A}
\end{eqnarray}
The magnetic field can be obtained from the vector potential through ${\bf H} = \frac{1}{\mu_0 } \nabla \times {\bf A(r)}$. Due to the PEC termination at $z=0$, we consider the quadrupole moment of the image scatterer, which is located at ${\bf r'_s}= (\rho'_s , \phi'_s , z'_s) = (0,0,-z_s)$. As a result, we obtain the magnetic field of both the quadrupole and its image with respect to the flange
\begin{multline}
    {\bf H}_q({\bf r \mp rs}) = -\frac{i \omega \mu_0}{8 \pi} Q_{zz} (z-z') \frac{\rho - \rho' \cos \phi'}{ | {\bf r-r_s}|^3} \\
    \times e^{ik | {\bf r-r_s}|} \left( -k^2 -\frac{3ik}{ | {\bf r-r_s}|} +\frac{3}{| {\bf r-r_s}|^2 }\right) {\bf \hat{\phi}}
\label{Formulation : H quadrupole}
\end{multline}


with $| {\bf r-r_s}| = \sqrt{\rho ^2 +\rho'^2 -2\rho \rho ' \cos \phi' + (z-z')^2}$.

Following the same procedure (see Appendix \ref{AppendixA}) for the quadrupole magnetic field, we obtain the quadrupole's contribution to the terminal admittance
\begin{multline}
\frac{Y_q(0)}{Y_0} = \frac{G_q(0)}{Y_0}
- i \frac{B_q(0)}{Y_0}  \\
= -\frac{i\omega }{4\pi} \frac{\eta_c \log\left(\frac{b}{a}\right)}
{[\int_a^b \varepsilon(\rho) d\rho]^2}
z_s^2 \frac{\alpha_q}{1-\alpha_q G^E_{dip,zz}(2z_s) } 
\\ \times
\left[ \int_a^b \varepsilon (\rho) d\rho \frac{\rho^2  e^{ik\sqrt{\rho^2 +z_s ^2}} }{(\rho^2 +z_s ^2)^{\frac{3}{2}}}
   \left( k^2  + \frac{3ik}{ \sqrt{ \rho^2 +z_s^2}} - \frac{3}{ \rho^2 +z_s^2 }\right) \right]^2  
\label{Formulation : Y_q}
\end{multline}

Now, the total terminal admittance becomes 
\begin{eqnarray}
\frac{Y_{tot}(0)}{Y_0} &=& 
\frac{Y(0)}{Y_0 } + \frac{Y_p(0)}{Y_0} + \frac{Y_q(0)}{Y_0}
\label{Formulation : Y_tot_with_q}
\end{eqnarray}

\subsection{Possible Approximations}\label{sec:approximations}

Most commonly, the field in the aperture plane is approximated as a pure TEM mode where $\varepsilon(\rho) = 1\slash \rho$. This lets us get a closed form expression, for both the dipole and quadrupole contributions
\begin{multline}
\frac{Y_p(0)}{Y_0} 
= -\frac{i\omega }{2\pi} \frac{\eta_c }
{\log\left(\frac{b}{a}\right)}
\frac{\alpha_d}{1-\alpha_d G^E_{dip,zz}(2z_s) }\\
 \times
\left[ \frac{e^{ik R_b}}
{R_b} - \frac{e^{ikR_a}}
{R_a}
\right] ^2
\label{Formulation : Y_s TEM}
\end{multline}


\begin{multline}
\frac{Y_q(0)}{Y_0} 
= -\frac{i\omega }{4\pi} \frac{\eta_c }
{\log\left(\frac{b}{a}\right)}
z_s^2 \frac{\alpha_q}{1-\alpha_q G^E_{dip,zz}(2z_s) } 
 \\
\times \left[ \frac{e^{ikR_a}}
{R^2_a} \left(ik - \frac{1}
{R_a}\right) - \frac{e^{ikR_b}}
{R^2_b} \left(ik - \frac{1}
{R_b}\right)
\right] ^2.
\label{Formulation : Y_q TEM}
\end{multline}

Here, $R_{a}=\sqrt{z_s^2+a^2}$ and $R_{b}=\sqrt{z_s^2+b^2}$. 

This approximation is valid for the case of a fairly small aperture, where the contribution of the higher modes to the terminal admittance is negligible ($kb\ll 1$). Moreover, as we move further away from the aperture, this approximation becomes more accurate. This is due to the contribution of the higher modes of the aperture field to the radiated fields outside the probe decaying faster than the TEM mode, as they contain larger transverse wavenumebrs. This expression can be further simplified by using a quasi-static approximation for the aperture field, where $ k z_s \ll 1$. This would also cause the effects of radiation loss from the scatterers to be negligible, and therefore proper adjustments must also be made to $\alpha_d$ and $\alpha_q$ for consistency by removing the second terms in equations (\ref{Formulation : alpha approc}) and (\ref{Formulation : alpha_q}).

Due to the configuration, the radiation of the image of the scatterer in the presence of the conducting flange was considered, and accounted for through $G^E_{dip,zz}(2z_s)$. A study of the effect of the image source on the scatterer’s dipolar moment, revealed that $|\alpha G^E_{dip,zz}| \ll 1$, obtaining maximum value of 0.001 for our parameters regime. For example, for a scatterer with $r_s = 0.6$ mm, this contribution translates into a variation of approximately $0.5\%$ in the obtained admittance, whereas the quadrupole contribution's variation is approximately $30 \%$. Therefore, the image contribution of the dipole moment to the effective moment is negligible for our problem specifications. The expression for the terminal impedance can then be further simplified to an expression linear with $\alpha_d$, since we now take equation (\ref{Formulation :P_s effective}) to be ${\bf p_s^{eff}} \approx \alpha_d {\bf E}_{ap,z}({\bf r_s})$.
Similarly, the image contribution of the quadrupole moment to the effective moment was examined and found to be negligible.
These simplifications allow us to estimate the scatterer properties easily from the measured terminal admittance $\widetilde{Y}(0) $.

\section{Applications and Results}\label{Applications}

\begin{figure}[ht]
\begin{center}
\noindent
  \includegraphics[width=3.2in]{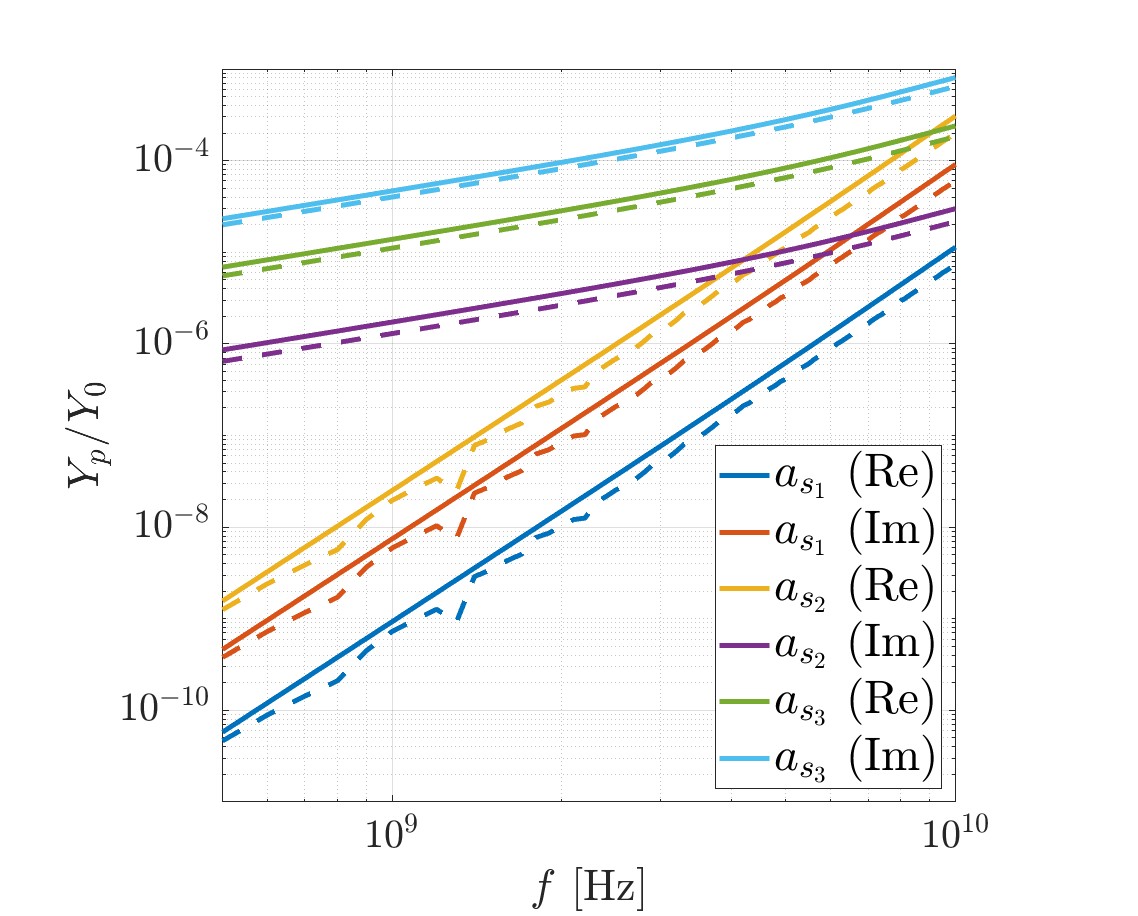}
\caption{Analytical modeling (solid line) and COMSOL full-wave simulations (dashed line) results for the scatterer terminal admittance when $\epsilon_c =1 , \epsilon_t = 15, \epsilon_s =20, a_s = 0.1, 0.2, 0.3 mm, z_s=1 mm$.}
\label{DifferentRadiiResults}
\end{center}
\end{figure}

Full-wave simulations were conducted using COMSOL Multiphysics \cite{COMSOLref}, to estimate the accuracy of the proposed model. The simulations followed the same set-up and parameters as in Fig. \ref{ConfigurationFig}.
Following the analysis in the previous section, let us define the contribution of the scatterer to the terminal admittance, using the following analytical expression
\begin{eqnarray}
\frac{Y_{s}(0)}{Y_0} &=& 
\frac{Y_p(0)}{Y_0} + \frac{Y_q(0)}{Y_0}
\label{Formulation : Y_s_simple}
\end{eqnarray}

where $\frac{Y_p(0)}{Y_0}$ and $\frac{Y_q(0)}{Y_0}$ are defined in equations (\ref{Formulation : Y_s TEM}) and (\ref{Formulation : Y_q TEM}), respectively.
Fig. \ref{DifferentRadiiResults} presents the real and imaginary parts of $Y_p(0)$, which is the dipole moment contribution of the small scatterer to the terminal admittance $Y_{tot}(0)$. We see a good agreement between the analytical modeling and full-wave simulations. Moreover, the frequency dependent behavior is very similar, which confirms the validity of our physical model.  

\begin{figure}[hbt]
\begin{center}
\noindent
  \includegraphics[width=\columnwidth]{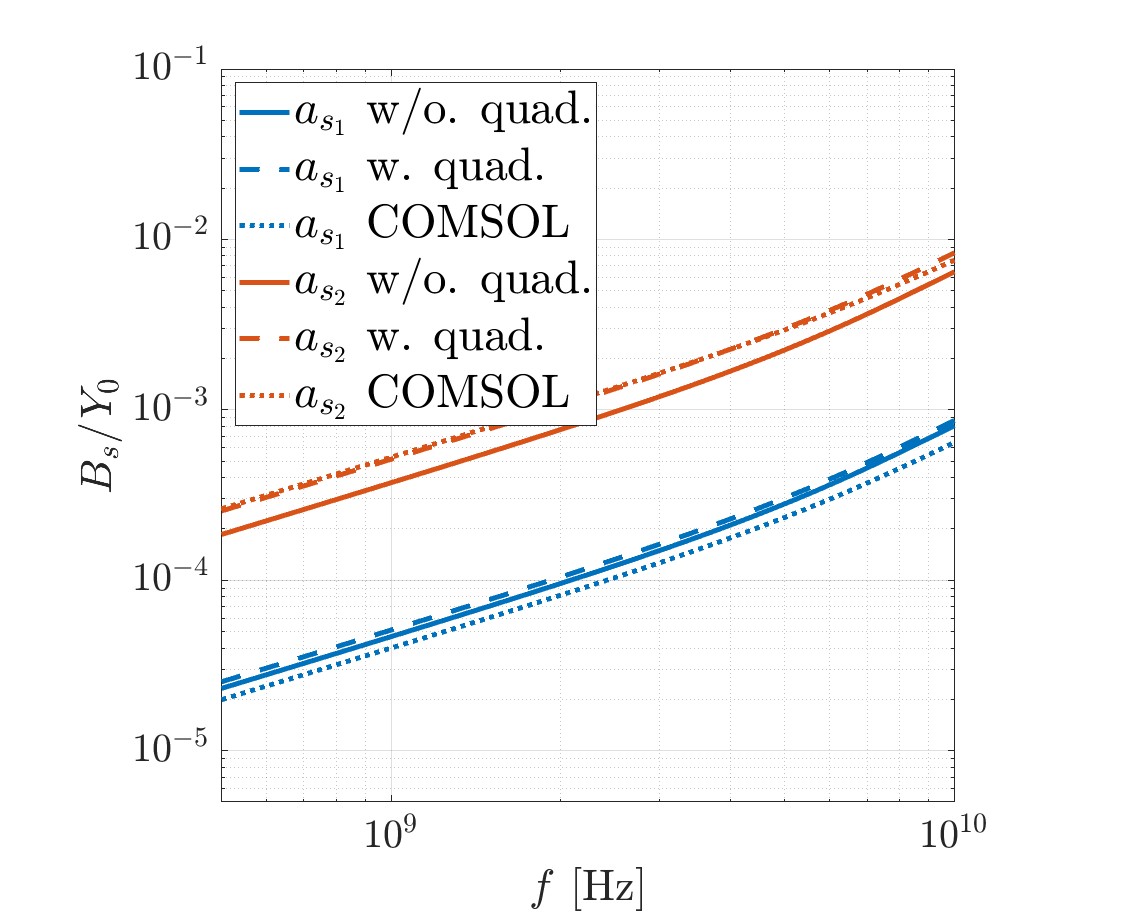}
\caption{ Analytical modeling results for the scatterer terminal susceptance, with and without the quadrupole contribution Vs. COMSOL full-wave simulations  ($\epsilon_c =1 , \epsilon_t = 15, \epsilon_s =20, a_{s_1} = 0.3$ mm, $a_{s_2} = 0.6$ mm, $z_s =1$ mm).}
\label{ResultQuad}
\end{center}
\end{figure}

In Fig. \ref{ResultQuad} we see a comparison of the contribution to the terminal susceptance, with and without the contribution of the quadrupole. Despite being deeply sub-wavelength, the quadrupole moment becomes significant for larger scatterers. Hence, by incorporating the quadrupole moment into the analytical model, the error is reduced. For example, for the larger scatterer with radius $a_s = 0.6$ mm, the error is reduced by $30\%-90\%$, when accounting for both the dipole and quadrupole contribution. The model becomes highly accurate when both contributions are used. However, there is still a residual error caused by the approximations made, higher modes of the aperture field, and small simulation errors.

\subsection{Parameters Extraction}\label{sec:ResultsParametersExtraction}



\begin{figure}[ht]
\begin{center}
\noindent
  \includegraphics[width=\columnwidth]{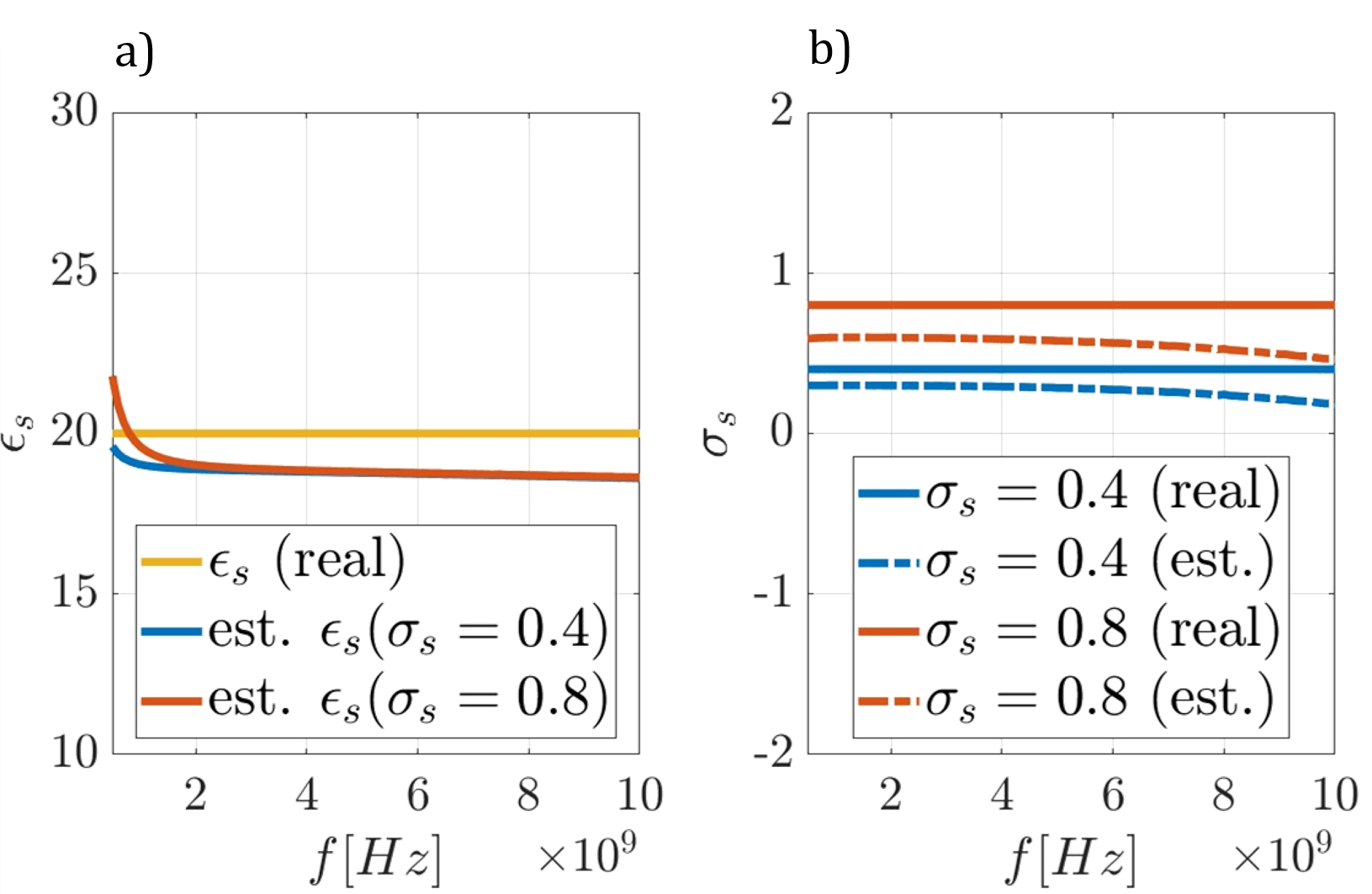}
\caption{$\epsilon_s$ estimations based on full-wave simulation results, where $\epsilon_t=15$ and $a_s = 0.2$ mm, using a probe with $2a = 0.65$ mm, $2b = 1.5$ mm, $\epsilon_c = 1$ }
\label{EstimationFig}
\end{center}
\end{figure}

In Fig. \ref{EstimationFig}, we extract the permittivity and conductivity of the scatterer from the full-wave simulation results using the approximate analytical formulas we obtained in (\ref{Formulation : Y_s TEM}). The complex permittivity is estimated based on only one simulated measurement, obtained using a single probe. We see good agreement between the estimated complex permittivity of the scatterer and the actual $\epsilon_s$ used for the simulation, for two different conductivity values $\sigma_s = 0.4,0.8$ [$S/ m$], with an error ranging between $2\%-7\%$ for the frequencies $f > 2$ GHz.

\subsection{Sensing Depth Estimation}\label{sec:ResultsSensingDepth}The sending depth and radius characterize the measurement in terms of the regions where a perturbation to the material under test will generate a variation in the measured reflection, above a certain threshold. In addition, it is specific for the scenario we consider, since it depends on the geometry and characteristics of the perturbations. The perturbation can take various forms. In the context of inhomogeneous medium sensing, one option is to examine a heterogeneity of infinite size, in the form of a layered medium. The layering can be in the axial or radial direction from the probe tip \cite{Lazebnik}.
Several other works have dealt with semi-infinite heterogeneities, analyzing the sensing radius or depth \cite{Emily_2018,Emily_La_Gioia,Meaney}. 
However, in this work we aim to examine how a finite heterogeneity affects the probe terminal admittance and, as a result, the extracted permittivity, with the goal of detecting and characterizing the heterogeneity using the OECP measurements. As we have shown in the previous section, the extraction of the dielectric properties of the heterogeneity depends on the size, location, properties of the background material, and the perturbations within the sensing volume of the probe \cite{Farshkaran,Brace}. Therefore, in the case of a non-homogeneous test sample, it is crucial to quantify the sensing depth of the probe accurately, to establish a common ground on whether a selected probe is suitable for the expected properties of the measured heterogeneity. In \cite{Brace} a comprehensive study was conducted based on numerical simulations. 

\begin{figure}[ht]
\begin{center}
\noindent
\includegraphics[width=\columnwidth]{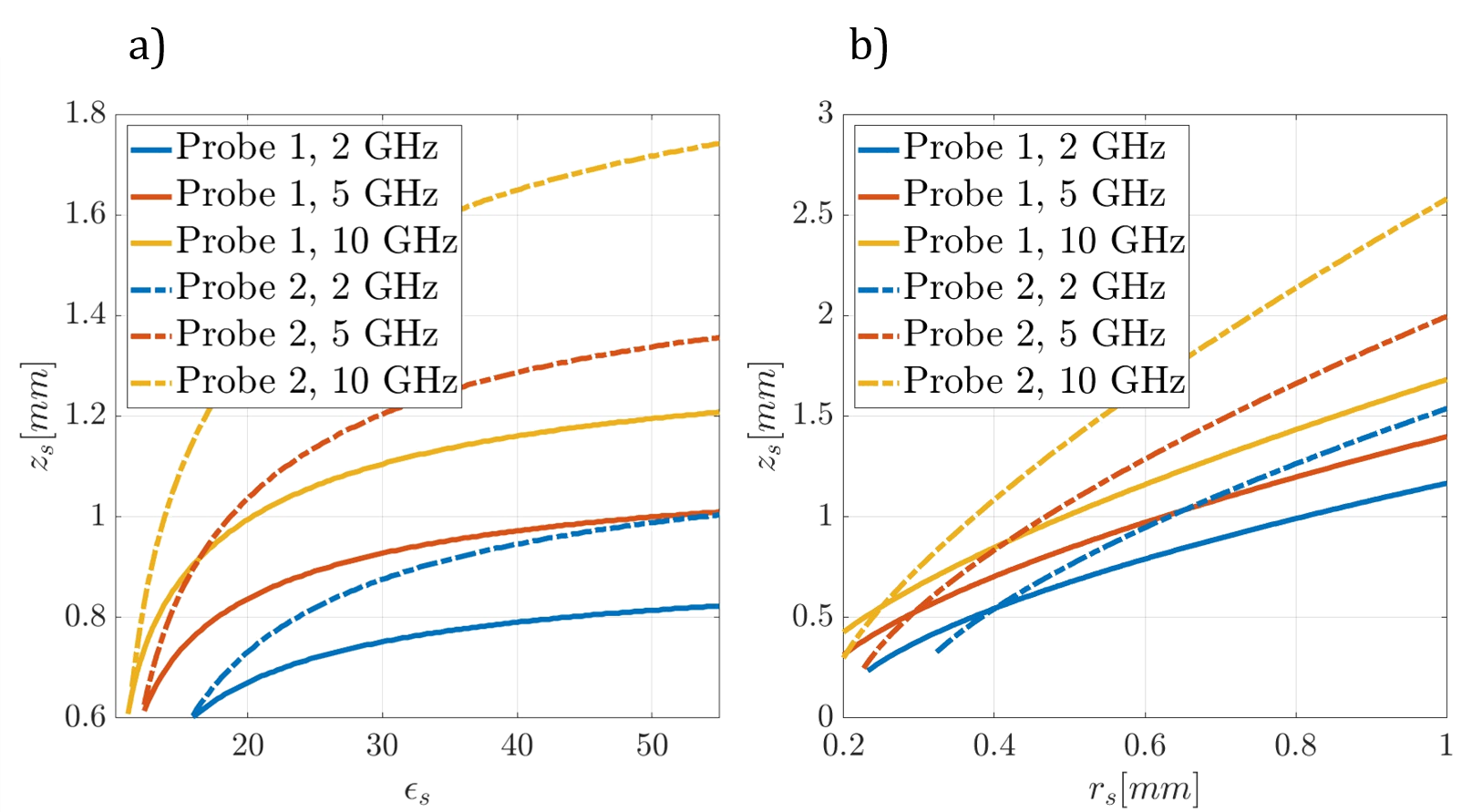}
\caption{Calculated sensing depth $z_s$ as a function of the scatterer (a) permittivity ($a_s = 0.6$ mm) (b) radius $(\epsilon_s = 40)$, where $\epsilon_t = 10, f = 2,5,10 $ GHz.}
\label{ZsVsEpsilonandRadius}
\end{center}
\end{figure}

Here, we suggest a simple and straightforward approach based on the derived analytical model to estimate the sensing depth of the probe. To this end, we first introduce the definition of the sensing depth. Using this definition, the sensing depth is defined as the distance from the interface at which a certain perturbation to the background medium generates a variation of $S_{11}$ above a certain threshold. In this case, an isotropic scatterer is inserted into an otherwise homogeneous tissue, with the measured reflection coefficient at the interface given by 
\begin{eqnarray}
S_{11} &=&
\frac{1- \frac{Y_{tot}(0)}{Y_0} }{1+ \frac{Y_{tot}(0)}{Y_0}}
\label{Results : S_11}
\end{eqnarray}

where $Y_{tot}(0)\backslash Y_0$ is given in (\ref{Formulation : Y_tot_with_q}).
As an example, a $2\%$ threshold (corresponding with a $5\% - 10\%$ deviation in the measured permittivity) was chosen, rendering our requirement for a viable measurement
\begin{eqnarray}
\left|\Delta S_{11}\right| = 
\left|S_{11_{w/s}} - S_{11_{wo/s}}\right| = 0.02
\label{Results : Threshold}
\end{eqnarray}

where the subscript $(\_)_{w/s}$ stands for “with scatterer” and $(\_)_{wo/s}$ stands for “without scatterer”. In Fig. \ref{ZsVsEpsilonandRadius}(a) we present the sensing depth as a function of the scatterer's permittivity.  In Fig. \ref{ZsVsEpsilonandRadius}(b) we present the sensing depth as a function of the scatterer's radius. 
The results refer to two different probes, with the following parameters - probe 1 with $2a = 0.65$ mm, $2b = 1.5$ mm, $\epsilon_c = 1$ and probe 2 with $2a = 0.93$ mm, $2b = 3.5$ mm, $\epsilon_c = 2.53$. As we can see from Fig. \ref{ZsVsEpsilonandRadius}(a), as the contrast between $\epsilon_t$ and $\epsilon_s$ increases, the scatterer's contribution becomes more dominant, and consequently, the sensing depth increases as well. This conclusion is consistent with the results in \cite{Emily_2018,Emily_La_Gioia}, though they focus on the sensing radius, rather than the sensing depth. However, due to the nonlinear dependence of the polarizability on the $\epsilon$ contrast between the inclusion and the environment, the sensing depth saturates at a value determined by the given scatterer radius $r_s$, the chosen probe and the threshold. Any inclusion within the background material will be considered "undetectable" beyond this depth. In Fig. \ref{ZsVsEpsilonandRadius}(b) we observe that the sensing depth increases, as the scatterer radius increases, since the scatterer's contribution becomes more dominant. Moreover, we are able to increase the sensing depth when using a larger probe due to higher field penetration into the measured medium. These results are in agreement with the cases studied in \cite{Brace,Farshkaran}, and using the analytical expressions, this inquiry can be extended to other various cases without any computational effort.

%

%
We formulate an analytical approximate lower boundary for the sensing depth, using the a quasi-static approximation for the background contribution $Y_t$, where $ k z_s \ll 1$, and assuming that the contribution of the scatterer is relatively small, i.e. $Y_p / Y_0 \ll 1$.
\begin{multline}
\left|\Delta S_{11}\right| \approx \\ 
\Bigg|  \frac{i\omega }{2\pi} \frac{\eta_c }
{\log\left(\frac{b}{a}\right)}
\frac{\alpha_d}{\left(1+ Y(0)/Y_0 \right)^2}
\left[ \frac{1}
{R_b} - \frac{1}
{R_a}
\right] ^2 \Bigg| = 0.02 
\label{Results : LowerBoundaryZs}
\end{multline}

\begin{figure}[ht]
\begin{center}
\noindent
\includegraphics[width=\columnwidth]{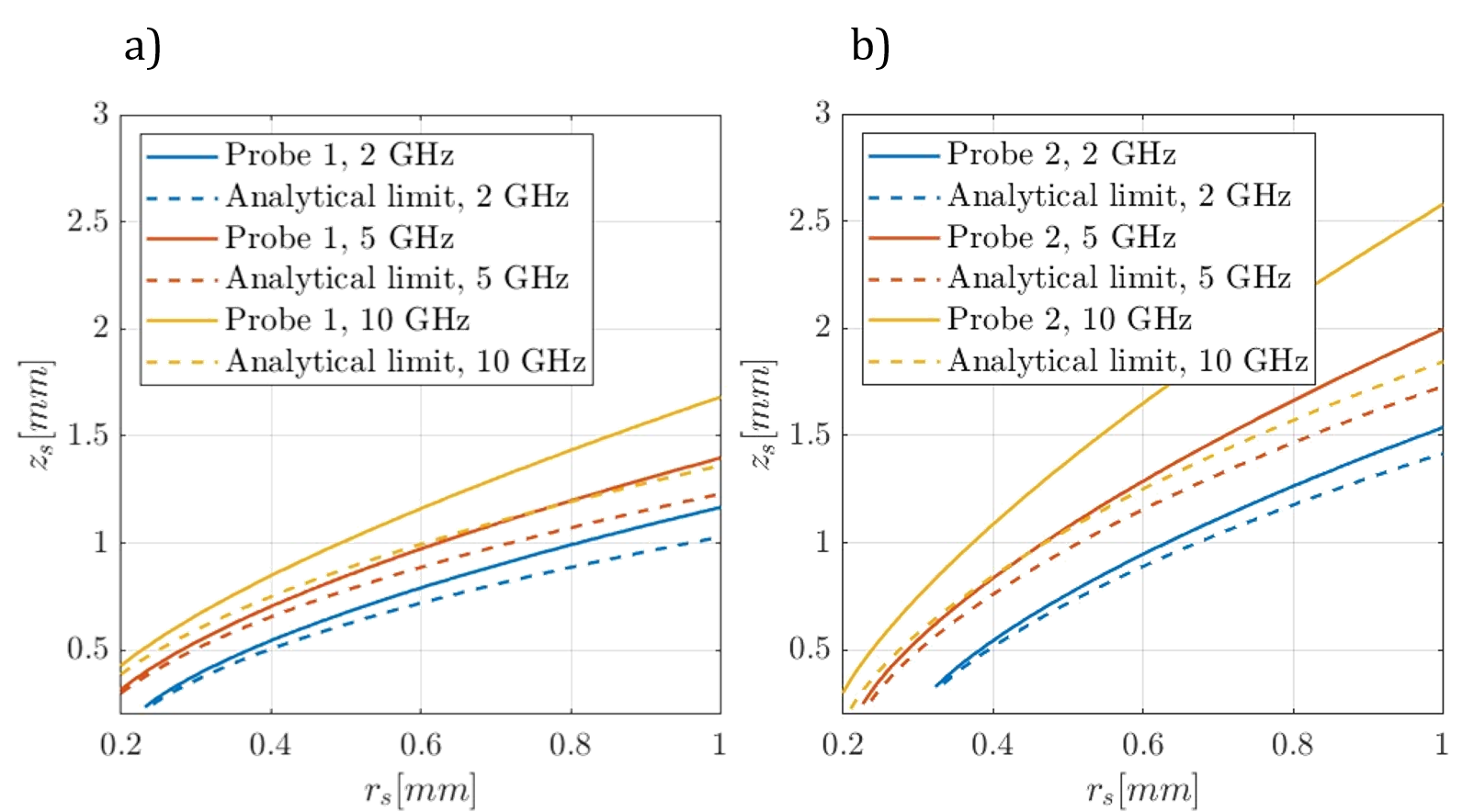}
\caption{Calculated sensing depth $z_s$ and $z_s$ analytical limit as a function of the scatterer radius $ \epsilon_t = 10, \epsilon_s = 40, f = 2,5,10 $ GHz, for $1.5$ mm probe with $2a = 0.65$ mm, $2b = 1.5$ mm, $\epsilon_c = 1$.}
\label{ZsVsRswLimitBothProbes}
\end{center}
\end{figure}

In Fig. \ref{ZsVsRswLimitBothProbes}(a) and \ref{ZsVsRswLimitBothProbes}(b) we present the comparison between the obtained sensing depth and the approximated lower boundary for the sensing depth, as a function of the scatterer's radius, for probe 1 and probe 2, respectively. It can be observed that, as expected, the sensing depth will increase when using a higher frequencies or larger probes. Experiments presented in \cite{Meaney} are in agreement with these results, and demonstrate the strong correlation between sensing depth and probe dimensions. Moreover, the larger the perturbation (detected scatterer) is, the easier it will be to detect, which also translates into a higher sensing depth. Comparing the lower boundary with the actual sensing depth, it is evident that overall behavior is similar, and as expected converges when the scatterer is very small. Using this bound can give us a good sense of the expected sensing depth, requiring only relatively simple understanding of the test sample. 

\section{Conclusion}\label{Conc}
In this study, we present an analytical model for the contribution of an isotropic scatterer to the measured terminal admittance, when using the OECP measurement technique. Explicit and simplified expressions for the excited dipole and quadrupole moments were derived, which are used to obtain closed-form expressions for the perturbed terminal admittance.
We then compared our analytical results with full-wave numerical simulations, showing excellent agreement. The derived model was utilized to obtain two key results: first, to estimate the scatterer complex permittivity, and second, to demonstrate a new method for estimating sensing depth. With its accuracy and strong basis in scattering theory, the derived model provides a significant glimpse into characterization of heterogeneous tissues' dielectric properties using OECP, and can be used as a guideline to model the effect of other types of inhomogeneities.

\appendices
\section{Deriving the Terminal Admittance Using a Variational Principle}\label{AppendixA}

The purpose of this section is to formulate a variational principle for the dipole and quadrupole contributions - $\frac{Y_p(0)}{Y_0}$ and $\frac{Y_q(0)}{Y_0}$, and find a representation of this quantity, using the magnetic field of the scatterer in the half space $z>0$, which was calculated before, and given in eq. (\ref{Formulation : H sphere}).

Across the aperture of the OECP, the magnetic field $H_\phi (\rho,z)$ must be continuous, i.e., $H_\phi^{(-)} (\rho,0)= H_\phi^{(+)} (\rho,0)$, where $H_\phi^{(-)} (\rho,z), H_\phi^{(+)} (\rho,z)$ are the magnetic fields inside the coaxial probe and in the half space given by $ z>0$, respectively. 

Using $H_\phi^{(-)} (\rho,z)$ given in \cite{Papas}, and considering the contribution of the scatterer's dipole moment to the external field, we obtain
\begin{multline}
    \frac{I(0)}{2\pi\rho} +
i\omega\epsilon_{c}\int^b_a \varepsilon(\rho')\rho'd\rho'
\sum_{n=1}^{\infty} \frac{R_n(\rho) R_n(\rho')}
{(\lambda _n^2 -k_{c}^2)^\frac{1}{2}} = \\
H_\phi^{(+)} (\rho,0) + H_{s_\phi} (\rho, 0 , \rho_s, z_s)+ H_{s_\phi} (\rho, 0 , \rho_s, -z_s) 
\label{AppendixA : ManFieldContinuity}
\end{multline}

Substituting the background material magnetic field and the scatterer's dipole moment magnetic field ${\bf H}_s( \rho, 0 , \rho_s, \mp zs) $ given by eq. (\ref{Formulation : H sphere}) we obtain

\begin{multline}
    \frac{I(0)}{2\pi\rho} +
i\omega\epsilon_{c}\int^b_a \varepsilon(\rho')\rho'd\rho'
\sum_{n=1}^{\infty} \frac{R_n(\rho) R_n(\rho')}
{(\lambda _n^2 -k_{c}^2)^\frac{1}{2}} = \\
- \frac{i\omega \epsilon_{t}}{2\pi} \int_a^b \varepsilon(\rho')\rho'd\rho'
 \int^{2\pi}_{0} d\phi \cos \phi 
\frac{e^{i k_{t}(\rho^2 +\rho'^2 -2\rho\rho'\cos\phi)^{\frac{1}{2}}}}
{(\rho^2 +\rho'^2 -2\rho\rho'\cos\phi)^{\frac{1}{2}}} \\
+ ( {\bf G}^H_{s} (\rho , 0, \rho_s ,z_s) + {\bf G}^H_{s} (\rho , 0, \rho_s , - z_s))
p_s  \frac{e^{ikr}}{4 \pi r}\\ 
\times  
\int_a^b \rho'^2 \varepsilon (\rho ')
\frac{ e^{ik\sqrt{\rho'^2 +z_s ^2}} }{\rho'^2 +z_s ^2} 
\left( -ik + \frac{1}{\sqrt{\rho'^2 +z_s ^2}}\right) d\rho'
\label{AppendixA : 2.35full}
\end{multline}

for the region $a \leq \rho \leq b$. 
Now, the following steps are taken:
\begin{enumerate}
    \item Since we are using the TEM approximation for the coaxial probe aperture field, we will omit the terms related to the higher modes.
    \item ${\bf G}^H_{s} (\rho , 0, \rho_s ,z_s) , {\bf G}^H_{s} (\rho , 0, \rho_s , - z_s))$, given in eq. (\ref{Formulation : G_H in cyl. Cor.}), are substituted in the equation.
    \item  The integral equation is multiplied by $\rho \varepsilon(\rho)$ and integrated from $\rho =a$ to $\rho =b$.
    \item The resultant equation is then  multiplied by 
    \begin{eqnarray}
    \nonumber
    \frac{({\mu_{c}}/ \epsilon_{c})^\frac{1}{2} \log (b/a)}{  [\int_a^b \varepsilon(\rho)d\rho]^2} &=& \frac{\eta_c \log (b/a)}{[\int_a^b \varepsilon(\rho)d\rho]^2}
    \label{AppendixA : Mult}
    \end{eqnarray}
    \item We recall the characteristic admittance of the probe, given by
    \begin{eqnarray}
    Y_0 &=& \frac{2\pi}{({\mu_{c}}/ \epsilon_{c})^\frac{1}{2}  \log (b/a) }
    \label{AppendixA : Y_0}
    \end{eqnarray}
    and re-arrange the equation accordingly.
\end{enumerate}

Finally, we derive an expression for the total terminal admittance in the presence of the scatterer
\begin{multline}
\frac{Y(0)}{Y_0} = \\
- \frac{ik_{t}
\sqrt{\frac{\epsilon_{t}}{\epsilon_{c}}}
}{2\pi \log(b/a)} \int_a^b \varepsilon(\rho)\rho d\rho
\int_a^b \varepsilon(\rho')\rho'd\rho'
\\ \times \int^{2\pi}_{0} d\phi \cos \phi 
\frac{e^{i k(\rho^2 +\rho'^2 -2\rho\rho'\cos\phi)^{\frac{1}{2}}}}
{(\rho^2 +\rho'^2 -2\rho\rho'\cos\phi)^{\frac{1}{2}}} 
\\ -\frac{i\omega }{2\pi} \frac{\eta_c \log\left(\frac{b}{a}\right)}
{[\int_a^b \varepsilon(\rho) d\rho]^2}
\frac{\alpha}{1-\alpha G^E_{dip,zz}(2z_s) }\\
 \times
 \left[ \int_a^b \varepsilon (\rho) d\rho \frac{\rho^2}{\rho^2 +z_s ^2}
 e^{ik\sqrt{\rho^2 +z_s ^2}} 
 \left( ik - \frac{1}{\sqrt{\rho^2 +z_s ^2}}\right) \right]^2  
\label{AppendixA : Y_t+Y_s}
\end{multline}

Since the result is the sum of contributions from the background material and the scatterer itself, the scatterer's dipole moment contribution can now be determined
\begin{multline}
\frac{Y_p(0)}{Y_0} =-\frac{i\omega }{2\pi} \frac{\eta_c \log\left(\frac{b}{a}\right)}
{[\int_a^b \varepsilon(\rho) d\rho]^2}
\frac{\alpha}{1-\alpha G^E_{dip,zz}(2z_s) }\\
 \times
 \left[ \int_a^b \varepsilon (\rho) d\rho \frac{\rho^2}{\rho^2 +z_s ^2}
 e^{ik\sqrt{\rho^2 +z_s ^2}} 
 \left( ik - \frac{1}{\sqrt{\rho^2 +z_s ^2}}\right) \right]^2  
\label{AppendixA : Y_s}
\end{multline}

When considering the secondary contribution, arising due to the quadrupole moment, equation (\ref{AppendixA : ManFieldContinuity}) becomes
\begin{multline}
    \frac{I(0)}{2\pi\rho} +
i\omega\epsilon_{c}\int^b_a \varepsilon(\rho')\rho'd\rho'
\sum_{n=1}^{\infty} \frac{R_n(\rho) R_n(\rho')}
{(\lambda _n^2 -k_{c}^2)^\frac{1}{2}} = \\
H_\phi^{(+)} (\rho,0) + H_{s_\phi} (\rho, 0 , \rho_s, z_s)+ H_{s_\phi} (\rho, 0 , \rho_s, -z_s) \\
 + H_{q_\phi} (\rho, 0 , \rho_s, z_s)+ H_{q_\phi} (\rho, 0 , \rho_s, -z_s) 
\label{AppendixA : ManFieldContinuityWQuad}
\end{multline}

where ${\bf H}_q( \rho, z , \rho_s, \mp zs) $ is the magnetic field due to the quadrupole moment of the scatterer and its image. Following the same procedure taken for the dipole moment, we derive the scatterer's quadrupole moment contribution to the terminal admittance

\begin{multline}
\frac{Y_q(0)}{Y_0} =
-\frac{i\omega }{4\pi} \frac{\eta_c \log\left(\frac{b}{a}\right)}
{[\int_a^b \varepsilon(\rho) d\rho]^2}
z_s^2 \alpha_q 
\\ \times
\left[ \int_a^b \varepsilon (\rho) d\rho \frac{\rho^2  e^{ik\sqrt{\rho^2 +z_s ^2}} }{(\rho^2 +z_s ^2)^{\frac{3}{2}}}
   \left( k^2  + \frac{3ik}{ \sqrt{ \rho^2 +z_s^2}} - \frac{3}{ \rho^2 +z_s^2 }\right) \right]^2  
\label{AppendixA : Y_q}
\end{multline}
\bibliographystyle{ieeetr}
\bibliography{OpenCoaxBib.bib}





\end{document}